\documentclass[showpacs,prc,twocolumn,floatfix,aps]{revtex4-1}
\usepackage{amssymb} 
\usepackage{amsmath} 
\usepackage{graphicx}
\usepackage{bm}
\usepackage{multirow}
\usepackage{color}
\graphicspath{{New_Figures/}}

\begin{document}
\title{Microscopic self-energy calculations and dispersive optical-model potentials}
\date{\today}
\author{S. J. Waldecker$^{1}$}
\author{C. Barbieri$^{2,3}$}
\altaffiliation{Permanent address: University of Surrey, Guildford, UK}
\email{C.Barbieri@surrey.ac.uk}
\author{W. H. Dickhoff$^{1}$}
\affiliation{$^{1}$Departments of Physics, Washington University, St.~Louis, Missouri 63130, USA}
\affiliation{$^{2}$Department of Physics, Faculty of Engineering and Physical Sciences, University of Surrey, Guildford, Surrey GU2 7XH, United Kingdom}
\affiliation{$^{3}$Theoretical Nuclear Physics Laboratory, RIKEN Nishina Center, 2-1 Hirosawa, Wako, Saitama 351-0198, Japan}
\begin{abstract}
Nucleon self-energies for  ${}^{40,48,60}$Ca isotopes are generated with the microscopic Faddeev-random-phase approximation (FRPA).
These self-energies are compared with potentials from the dispersive optical model (DOM) that were obtained from fitting elastic-scattering and bound-state data for ${}^{40,48}$Ca.
The \textit{ab initio}  FRPA is capable of explaining many features of the empirical DOM potentials including their nucleon asymmetry dependence.
The comparison furthermore provides several suggestions to improve the functional form of the DOM potentials, including among others the exploration of parity and angular momentum dependence.
The non-locality of the FRPA imaginary self-energy, illustrated by a substantial orbital angular momentum dependence, suggests that future DOM fits should consider this feature explicitly.  
The roles of the nucleon-nucleon tensor force and charge-exchange component in generating the asymmetry dependence of the FPRA self-energies are explored. 
The global features of the FRPA self-energies are not strongly dependent on the choice of realistic nucleon-nucleon interaction. 
\end{abstract}
\maketitle

\section{Introduction}
\label{Sec:introduction}
The properties of nucleons propagating in the nucleus exhibit characteristic deviations from the naive shell-model picture. 
At positive energies, corresponding to the domain of elastic scattering, there is unambiguous evidence that the potential (or self-energy) that a nucleon experiences is absorptive~\cite{Becchetti69,Varner91,Koning03}.
This simple observation has important implications, since it shows that nuclear states cannot be interpreted only in terms of a simple shell-model potential that is real and independent of energy.
The importance of the dynamic aspects of the nuclear shell model was recognized in Ref.~\cite{Mahaux85}.
The link between optical potentials and the traditional bound-state shell model was explored by Mahaux and Sartor~\cite{Mahaux86} and extensively reviewed in Ref.~\cite{Mahaux91}.
These authors realized that information on nucleon propagation at positive energy influences the properties of the real nuclear potential at negative energy, since the nucleon self-energy obeys a dispersion relation that links the real part to its imaginary part at all energies (see \textit{e.g.} Ref.~\cite{Dickhoff08}).
Mahaux and Sartor exploited standard representations of the imaginary part of the optical potential in terms of volume and surface contributions. They further assumed that the behavior of the imaginary potential was similar near both sides of the Fermi energy and used a subtracted form of the dispersion relation to obtain the corresponding real part.
By performing this subtraction at the Fermi energy, only the additional knowledge of the real potential at that energy is required. The resulting optical potential is now called the dispersive optical model (DOM).

Recent applications of the DOM have concentrated on the nucleon asymmetry dependence by simultaneously fitting data pertaining to different calcium isotopes~\cite{charity06,charity07} and to spherical isotopes up to Tin and $^{208}$Pb~\cite{Mueller11}.
Such an analysis can be utilized to predict properties of isotopes with larger nucleon asymmetry by extrapolating the DOM potentials.
Such data-driven extrapolations present a reliable strategy to approach and predict properties of isotopes towards the respective drip lines, since they can be tested by performing corresponding experiments.
An important feature extracted from this analysis is the increase in surface absorption of protons for increasing nucleon asymmetry.
While this trend is unambiguous, there is no clear understanding of the underlying dynamics responsible for it.
A much weaker and opposite trend was inferred for neutrons.
It is therefore useful to study the nucleon self-energy---which is the microscopic counterpart of the DOM---to clarify this behavior and provide a deeper understanding of the DOM potentials.

The Green's function method is ideally suited to pursue a microscopic understanding of the nucleon self-energy at both positive and negative energies~\cite{Dickhoff04}.
The most sophisticated implementation of the Green's function method considers the role of long-range or low-energy correlations in which nucleons couple to low-lying collective states and giant resonances.
This is accomplished by using the random phase approximation (RPA) to calculate phonons of particle-particle (hole-hole) and particle-hole type. These are then summed to all orders in a Faddeev summation for both two-particle--one-hole (2p1h) and two-hole--one-particle (2h1p) propagation. This approach is referred to as Faddeev random phase approximation (FRPA)~\cite{Barbieri07,Barb:1}.
This method is size extensive and has been successfully benchmarked for soft interactions in purely \textit{ab-initio} calculations for ${}^4$He~\cite{Barbieri10var}, giving results of comparable accuracy to coupled-cluster theory.

The FRPA was originally developed to describe the self-energy of the double closed-shell nucleus ${}^{16}$O~\cite{Barb:1,Barb:2}.
The method has later been applied to atoms and molecules~\cite{Barbieri07,Degroote11} and recently to ${}^{56}$Ni~\cite{Barbieri09a} and ${}^{48}$Ca~\cite{Barbieri09b}. 
The \textit{ab initio} results of Ref.~\cite{Barbieri09b} are in good agreement with $(e,e'p)$ data for spectroscopic factors from Ref.~\cite{Kramer01} and also show that the configuration space needed for the incorporation of long-range (surface) correlations is much larger than the space that can be utilized in large-scale shell-model diagonalizations.
In Ref.~\cite{Barbieri05}, the FRPA was employed to calculate proton scattering on $^{16}$O and obtain results for phase shifts and low-lying states in $^{17}$F. However, the properties of the self-energy at larger scattering energies which are now of great interest for the developments of DOM potentials was not addressed.
 In particular, one may expect to extract useful information regarding the functional form of the DOM from a study of the self-energy for a sequence of calcium isotopes. It is the purpose of the present work to close this gap.
We have chosen in addition to ${}^{40}$Ca and ${}^{48}$Ca also to include ${}^{60}$Ca, since the latter isotope was studied with a DOM extrapolation in Refs.~\cite{charity06,charity07}.
Some preliminary results of these FRPA calculations for spectroscopic factors were reported in Ref.~\cite{Barbieri10var} but the emphasis in the present work is on the properties of the microscopically calculated self-energies.
The resulting analysis is intended to provide a microscopic underpinning of the qualitative features of empirical optical potentials.
Additional information concerning the degree and form of the non-locality of both the real and imaginary parts of the self-energy will also be addressed because it is of importance to assess the current local implementations of the DOM method.

In Sec.~\ref{sec:self} we introduce some of the basic properties for the analysis of the self-energy.
The ingredients of the FRPA calculation are presented in Sec.~\ref{sec:FRPA}.
The choice of model space and realistic nucleon-nucleon (NN) interaction are discussed in Sec.~\ref{sec:space}.
We present our results in Sec.~\ref{sec:results} and finally draw conclusions in Sec.~\ref{sec:con}.

\section{Formalism}
\label{sec:formalism}

In the Lehmann representation, the one-body Green's function is given by
\begin{eqnarray}
 g_{\alpha \beta}(E) &=&
 \sum_n  \frac{ 
          \langle {\Psi^A_0}     \vert c_\alpha        \vert {\Psi^{A+1}_n} \rangle
          \langle {\Psi^{A+1}_n} \vert c^{\dag}_\beta  \vert {\Psi^A_0} \rangle
              }{E - (E^{A+1}_n - E^A_0) + i \eta }
\nonumber \\
&& +
 \sum_k \frac{
          \langle {\Psi^A_0}     \vert c^{\dag}_\beta  \vert {\Psi^{A-1}_k} \rangle
          \langle {\Psi^{A-1}_k} \vert c_\alpha        \vert {\Psi^A_0} \rangle
             }{E - (E^A_0 - E^{A-1}_k) - i \eta } \; ,
\label{eq:gsp}
\end{eqnarray}
where $\alpha$, $\beta$, ..., label a complete orthonormal basis set and
$c_\alpha$~($c^\dag_\beta$) are the corresponding second quantization 
destruction (creation) operators.
In these definitions, $\vert\Psi^{A+1}_n\rangle$, $\vert\Psi^{A-1}_k\rangle$ are the eigenstates, and $E^{A+1}_n$, $E^{A-1}_k$ the eigenenergies of the ($A\pm1$)-nucleon isotope.
The structure of Eq.~(\ref{eq:gsp}) is particularly useful for our purposes.
At positive energies, the residues of the first term, $\langle {\Psi^{A+1}_n} \vert c^{\dag}_\alpha \vert {\Psi^A_0} \rangle$, contain the scattering wave functions for the elastic collision of a nucleon off the $|\Psi^A_0\rangle$ ground state,  while at negative energies they give information on final states of the nucleon capture process. Consequently, the second term has poles below the Fermi energy ($E_F$) which carry information about the removal of a nucleon and therefore clarify the structure of the target state $|\Psi^A_0\rangle$ itself. Green's function theory provides a natural framework for describing physics both above and below the Fermi surface in a consistent manner.

The propagator~(\ref{eq:gsp}) can be obtained as a solution of the Dyson equation,
\begin{equation}
 g_{\alpha \beta}(E) ~=~  g^{(0)}_{\alpha \beta}(E) ~+~
   \sum_{\gamma \delta} 
    g^{(0)}_{\alpha \gamma}(E)  \,
     \Sigma^\star_{\gamma \delta}(E) \,  g_{\delta \beta}(E) \;  ,
\label{eq:Dyson}
\end{equation}
in which $g^{(0)}(E)$ is the propagator for a free nucleon (moving only with its kinetic energy). $\Sigma^\star(E)$ is the irreducible self-energy and represents the interaction of the projectile (ejectile) with the target nucleus.
Feshbach, developed a formal microscopic theory for the optical potential already in Ref.~\cite{Feshbach58, Feshbach62} by projecting the many-body Hamiltonian on the subspace of scattering states. It has been proven that if Feshbach's
theory is extended to a space including states both above and below the Fermi surface, the resulting optical potential
is exactly the irreducible self-energy $\Sigma^\star(E)$~\cite{Capuzzi96}~(see also Ref.~\cite{Bell59} and Ref.~\cite{Escher02} for a shorter demonstration).

The above equivalence with the microscopic optical potential is fundamental for the present study, since the available knowledge from calculations based on Green's function theory can be used to suggest improvements of optical models.
In particular, in the DOM, the dispersion relation obeyed by $\Sigma^\star(E)$ is used to reduce the number of parameters and to enforce the effects of causality.
Thus the DOM potentials can also be thought of as a representation of the nucleon self-energy.

\subsection{Self-Energy}
\label{sec:self}
For a $J = 0$ nucleus, all partial waves $(\ell , j, \tau )$ are decoupled, where $\ell$,$j$ label the orbital and total angular momentum and $\tau$ represents its isospin projection. The irreducible self-energy in coordinate space (for either a proton or a neutron) can be written in terms of the harmonic-oscillator basis used in the FRPA calculation, as follows:
\begin{eqnarray}
\Sigma^\star( \bm{x}, \bm{x^\prime}; E ) = \sum_{\ell j m_j \tau} {\cal I }_{\ell j m_j}( \Omega, \sigma ) \hspace{2.cm} \nonumber \\
\times \left[ \sum_{n_a, n_b} R_{n_a \ell}(r) \Sigma^\star_{ab}(E)R_{n_b \ell}( r^\prime )\right] ( {\cal I }_{\ell j m_j}( \Omega^\prime, \sigma^\prime ) )^* ,
\label{eq:selfr}
\end{eqnarray}
where $\bm{x} \equiv \bm{r}, \sigma, \tau$.  
The spin variable is represented by $\sigma$, $n$ is the principal quantum number of the  harmonic oscillator, and $a\equiv(n_a, \ell , j, \tau)$ (note that for a $J = 0$ nucleus the self-energy is independent of $m_j$).
The standard radial harmonic-oscillator function is denoted by $R_{n \ell}(r)$, while ${\cal I }_{\ell j m_j}( \Omega, \sigma )$ represents the $j$-coupled angular-spin function. 

We directly calculate the harmonic-oscillator projection of the self-energy, which can be written as
\begin{eqnarray}
\Sigma^\star_{ab}( E ) &=& \Sigma^{\infty}_{ab}(E) + \tilde{\Sigma}_{ab}(E) \nonumber  \\
&=& \Sigma^{\infty}_{ab}(E) + \sum_{r}{\frac{m_{a}^r( m_{b}^r )^*}{E -\varepsilon_r \pm i\eta}} .
\label{eq:selfHO}
\end{eqnarray}
The term with the tilde is the dynamic part of the self-energy due to long-range correlations calculated in the FRPA, and $\Sigma^{\infty}_{ab}(E)$ is the correlated Hartree-Fock term which acquires an energy dependence through the energy dependence of the $G$-matrix effective interaction (see below). $\Sigma^{\infty}_{ab}(E)$ is the sum of the strict correlated Hartree-Fock diagram (which is energy independent) and the dynamical contributions due to short-range interactions outside the chosen model space.
The self-energy can be further decomposed in a central ($0$) and a spin-orbit ($\ell s$) part according to
\begin{subequations}
\label{eq:ls}
\begin{eqnarray}
\Sigma^{\ell j_>} &=& \Sigma^{\ell}_{0}+\frac{\ell}{2}\Sigma^{\ell}_{\ell s} \label{eq:ls1} \, ,\\
\Sigma^{\ell j_<} &=& \Sigma^{\ell}_{0} - \frac{ \ell + 1}{2}\Sigma^{\ell}_{\ell s}  \, ,
\label{eq:ls2}
\end{eqnarray}
\end{subequations}
with $j_{>,<}\equiv \ell \pm \frac{1}{2}$.
The corresponding static terms are denoted by $\Sigma^{\infty, \ell}_0$ and $\Sigma^{\infty, \ell}_{\ell s}$, and the corresponding dynamic terms are denoted by $\tilde{\Sigma}^{\ell}_0$ and $\tilde{\Sigma}^{\ell}_{\ell s}$.

The FRPA calculation employs a discrete single-particle basis in a large model space which results in a substantial number of poles in the self-energy~(\ref{eq:selfHO}).
Since the goal is to compare with optical potentials at positive energy, it is appropriate to smooth out these contributions by employing a finite width for these poles.
We note that the optical potential was always intended to represent an average smooth behavior of the nucleon self-energy~\cite{Mahaux91}.
In addition, it makes physical sense to at least partly represent the escape width of the continuum states by this procedure.
Finally, further spreading of the intermediate states to more complicated states ($3p2h$ and higher excitations that are not included in the present calculation) can also be accounted for by this procedure.
Thus, before comparing to the DOM potentials, the dynamic part of the microscopic self-energy was smoothed out using a finite, energy-dependent width for the poles
\begin{equation}
\tilde{\Sigma}_{n_a, n_b}^{\ell j}(E) = \sum_{r}\frac{m_{n_a}^{r} m_{n_b}^{r}}{ E - \varepsilon_r \pm i\eta } \longrightarrow \sum_{r}\frac{m_{n_a}^{r} m_{n_b}^{r}}{ E - \varepsilon_r \pm i\Gamma(E) } \, .
\label{eq:width}
\end{equation}
Solving for the real and imaginary parts we obtain
\begin{eqnarray}
\lefteqn{
\tilde{\Sigma}_{n_a, n_b}^{\ell j}(E) = \sum_{r}\frac{(E - \varepsilon_r )}{ ( E - \varepsilon_r )^2 + [\Gamma(E)]^2 }m_{n_a}^{r} m_{n_b}^{r} 
 } & &	
 \label{eq:smooth} \\
&\qquad +& i\left[\theta( E_F - E )\sum_{h}\frac{\Gamma}{( E - \varepsilon_h )^2 + \Gamma(E)^2}m_{n_a}^{h}m_{n_b}^h  \right. \nonumber \\
&\qquad -& \left.  \theta( E - E_F )\sum_{p}\frac{\Gamma}{( E - \varepsilon_p )^2 + [\Gamma(E)]^2}m_{n_a}^{p}m_{n_b}^p\right] ,
\nonumber 
\end{eqnarray}
where, $r$ implies a sum over both particle and hole states, $h$ denotes a sum over the hole states only, and $p$ a sum over the particle states only. 
For the width, the following form was used~\cite{Brown81}:
$$
\Gamma( E ) = \frac{1}{\pi}\frac{ a \, (E- E_F)^2}{(E - E_F)^2 - b^2}
$$
with $a$=12~MeV and $b$=22.36~MeV.
This generates a narrow width near $E_F$ that increases as the energy moves away from the Fermi surface, in accordance with observations. 

In the DOM representation of the optical potential the self-energy is recast in the form of a  subtracted dispersion relation
\begin{equation}
\Sigma^\star_{ab}( E ) = \Sigma^{\infty}_{ab, \,S} + \tilde{\Sigma}_{ab}(E)_S, 
\label{eq:suba}
\end{equation}
where
\footnote{It is the (real) $\Sigma^\infty_{ab, \, S}$ and the imaginary part of $\tilde{\Sigma}_{ab}(E)_S$ that are 
parametrized in the DOM potential. \hbox{$Re \; \tilde{\Sigma}_{ab}(E)_S$} is then fixed by the subtracted dispersion relation.}
\begin{eqnarray}
       \Sigma^{\infty}_{ab \, S} &= &\Sigma^\star_{ab}(E_F) \, ,
\label{eq:SigSubHF} \\
\tilde{\Sigma}_{ab}(E)_S    &= &\Sigma^\star_{ab}(E) - \Sigma^\star_{ab}(E_F) \, .
\label{eq:SigSubDyn}
\end{eqnarray}		     
For the imaginary potential, this is the same as the above defined self-energies (\ref{eq:selfHO}) and it can therefore be directly compared to the DOM potential. For the real parts we will employ either the normal or the subtracted form in the following as appropriate.

\subsection{Volume Integrals}
In fitting optical potentials, it is usually found that volume integrals are better constrained by the experimental data~\cite{Mahaux91,Greenless68}. For this reason, they have been considered as a reliable measure of the total strength of a potential. For a non-local and $\ell$-dependent potential of the form~(\ref{eq:selfr}) it is convenient to consider separate integrals for each angular momentum component, $\Sigma^{\ell}_0(r, r^\prime)$ and $\Sigma^{\ell}_{\ell s}(r, r^\prime)$, which correspond to the square brackets in Eq.~(\ref{eq:selfr}) and decomposed according to~(\ref{eq:ls}).
Labeling the central real part of the optical potential with $V$, and the central imaginary part by $W$,  we calculate:
\begin{subequations}
\label{eq:intgs}
\begin{eqnarray}
J_W^\ell(E) = 4\pi\int{drr^2\int{dr^\prime r^{\prime 2} \text{Im } \Sigma^{\ell}_0(r, r^\prime ; E)}}
\label{eq:intgs_W} \\
J_V^\ell(E) = 4\pi\int{drr^2\int{dr^\prime r^{\prime 2} \text{Re } \Sigma^{\ell}_0(r, r^\prime ; E)}}
\label{eq:intgs_V}  .
\end{eqnarray}
\end{subequations}
We also employ the volume integral of the central real part at the Fermi energy denoted by $J_F^\ell = J_V^\ell(E_F)$,
and the corresponding averaged quantities
\begin{subequations}
\label{eq:intgs_avg}
\begin{eqnarray}
J_W^{avg}(E) = \frac{1}{N_{\{\ell\}}} \; \sum_{\ell \in \{\ell\}} J_W^\ell(E)
\label{eq:intgs_avg_W} \\
J_V^{avg}(E) = \frac{1}{N_{\{\ell\}}} \; \sum_{\ell \in \{\ell\}} J_W^\ell(E)  \, .
\label{eq:intgs_avg_V}
\end{eqnarray}
\end{subequations}
In Eqs.~(\ref{eq:intgs_avg}), $N_{\{\ell\}}$ is the number of partial waves included in the average and the sum runs over all values of $\ell$ except if otherwise indicated.
We also introduce the notation $J_F^{avg}=J_V^{avg}(E_F)$.

The correspondence between the above definitions and the volume integrals used for the (local) DOM potential in Refs.~\cite{charity06,charity07} can be obtained by casting a spherical local potential $U(r)$ into a non-local form $U(\bm r, \bm r^\prime ) = U(r) \delta(\bm r - \bm r^\prime)$. Expanding this in spherical harmonics gives
\begin{equation}
 U(\bm r, \bm r^\prime ) = \sum_{\ell m} U^\ell(r, r^\prime)Y^*_{\ell m}(\Omega^\prime)Y_{\ell m}(\Omega) \, ,
\end{equation}
with the $\ell$-projection
\begin{equation}
 U^\ell(r, r^\prime ) = \frac {U(r)}{r^2} \delta(r - r^\prime) \, ,
\end{equation}
which is actually angular-momentum independent.
The definition~(\ref{eq:intgs}) for the volume integrals lead to
\begin{eqnarray}
 J^\ell_U &=& 4\pi  \int{ dr\ r^2\int{dr^\prime r^{\prime 2} U^{\ell}(r, r^\prime)}} 
\\
      &=& 4\pi\int{ U(r) r^2dr} = \int{ U(r)\ d\bm{r}}  \, \hbox{, \hspace{.2in} for any $\ell$}
\nonumber
\end{eqnarray}
and reduces to the usual definition of volume integral for local potentials.
Thus, Eqs.~(\ref{eq:intgs}) and~(\ref{eq:intgs_avg}) can be directly compared to the corresponding integrals determined in previous studies of the DOM.

\subsection{Ingredients of the Faddeev-random-phase approximation}
\label{sec:FRPA}

\begin{figure}[b]
  \includegraphics[width=\columnwidth]{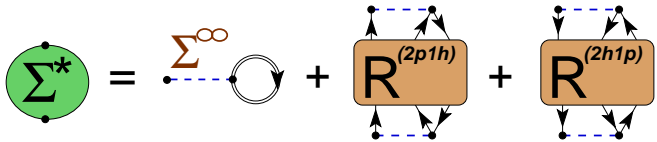}
  \caption{The self-energy $\Sigma^\star(E)$ separates exactly into
   a static (mean-field) term, $\Sigma^{\infty}$, and the polarization propagators
   $R^{(2p1h/2h1p)}(E)$ for the 2p1h/2h1p motion.  
    These $R(E)$ are expanded in terms of particle-vibration
   couplings as depicted below in Fig.~\ref{fig:faddex}.
    \label{fig:sigma_exp} }
\end{figure}

The self-energy is shown in terms of Feynman diagrams in Fig.~\ref{fig:sigma_exp}. 
The calculations are carried out in two steps by following the same procedure as in Ref.~\cite{Barbieri09a}, where further details can be found. 
First, a configuration space is selected that should be as large as possible to account for the treatment of nuclear collective motion.
We then account for the short-range part of a realistic NN interaction by directly calculating the two-body scattering for nucleons that propagate outside the model space. The result is the so-called $G$-matrix that must be employed as an energy-dependent effective interaction inside the chosen space. 
The contribution from ladder diagrams from outside the model space are also added to the calculated self-energy and result in an energy-dependent correction to $\Sigma^\infty_{ab}$~[see Eq.~(\ref{eq:selfHO})].
When the corresponding self-energy is calculated, this energy dependence enhances the reduction of the spectroscopic strength of occupied orbits by about 10\%. 
A similar depletion is also obtained in nuclear-matter calculations with realistic interactions~\cite{Dickhoff04} and confirmed by high-energy electron scattering data~\cite{Rohe04,Barbieri04b}.
The details of this partitioning procedure are presented in Ref.~\cite{Barbieri09a}. For the present discussion, it should be clear that this corresponds to calculating separately the contribution of propagators that lie outside the model space and then to add it to the final FRPA results. This does not introduce phenomenological parameters and the calculation should be regarded as a microscopic study based only on the original realistic interaction.

\begin{figure}
\includegraphics[width=\columnwidth,clip=true]{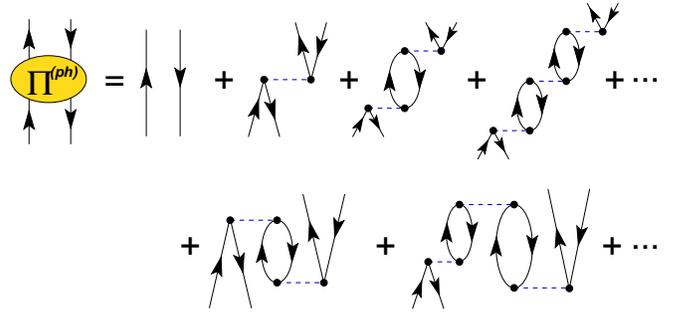}
\caption{
  Expansion of the ph propagator $\Pi(E)$ in a series of ring diagrams.
  The second line gives examples of time-inversion patterns that are generated by the RPA.
 A similar expansion, in terms of ladders diagrams, applies to $g^{II}(E)$. 
 The diagrams are time ordered, with time propagating upward.}
\label{fig:tda-rpa-eq}
\end{figure}

\begin{figure}[t]
\includegraphics[width=.30\textheight,clip=true]{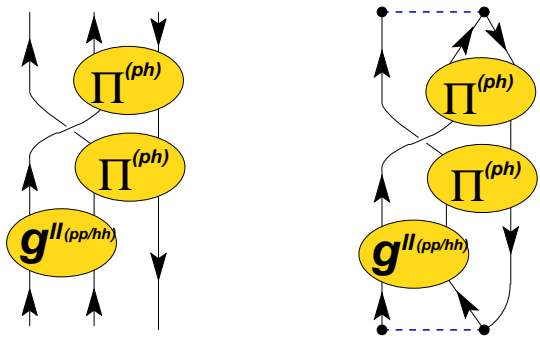}
\caption{ {\em Left:} Example of one of the diagrams
for $R^{(2p1h)}(E)$ that are summed to all orders by means of the
Faddeev method. Each of the ellipses represent an infinite sum of 
rings~[$\Pi(E)$] or ladders~[$g^{II}(E)$].
The diagrams included in $\Pi(E)$ are shown in Fig.~\ref{fig:tda-rpa-eq}
and $g^{II}(E)$ is the analogous for ladders~\cite{Barbieri07}.
 {\em Right:} The corresponding contribution to the self-energy obtained
from $R^{(2p1h)}(E)$ (compare to Fig.~\ref{fig:sigma_exp}).
 }
\label{fig:faddex}
\end{figure}

In addition to the influence of short-range (and tensor) correlations, it is essential to consider the role of long-range correlations in which nucleons couple to low-lying collective states and giant resonances. This is calculated in the second step inside the model space by employing the FRPA method.
The physics content of the FRPA is better summarized by looking at its diagrammatic expansion illustrated in Figs.~\ref{fig:tda-rpa-eq} and~\ref{fig:faddex}.
The basic ingredients are the particle-hole (ph) polarization propagator, $\Pi_{\alpha \beta , \gamma \delta}(E)$, that describes excited states of the $A$-nucleon system,
and the two-particle propagator, $g^{II}_{\alpha \beta , \gamma \delta}(E)$, that describes the propagation of two added/removed particles.
These propagators are calculated as summations of ring and ladder diagrams in the random-phase approximation (RPA). This allows for a proper description of collective excitations in the giant-resonance region when the model space is sufficiently large. The RPA induces time orderings as those shown in Fig.~\ref{fig:tda-rpa-eq} for the ph case and accounts for the presence of two-particle--two-hole and more complicated admixtures in the ground state, which are generated by correlations.
In FRPA, the $R^{(2p1h)}(E)$ and $R^{(2h1p)}(E)$ propagators that appear in Fig.~\ref{fig:sigma_exp} are obtained by recoupling $\Pi(E)$ and $g^{II}(E)$ to single-particle or hole states, as shown in  Fig.~\ref{fig:faddex}. This is done by solving the set of Faddeev equations detailed in Refs.~\cite{Barb:1,Barbieri07}.
Contributions from ph, particle-particle and hole-hole excitations in all possible partial waves are included in FRPA as this is required for a complete solution of the problem. Moreover, $R^{(2p1h)}(E)$ and $R^{(2h1p)}(E)$ also include energy-independent vertex corrections to ensure consistency with perturbation theory up to third-order to guarantee accurate results at the Fermi surface~\cite{Trofimov05}. We refer the reader to Ref.~\cite{Barbieri07} for more details.

The reference state employed in calculating the FRPA self-energy corresponds to a Slater determinant and is chosen to optimally approximate the fully correlated propagator~(\ref{eq:gsp}) near the Fermi energy. Once the self-energy is obtained, a new propagator is calculated by solving the Dyson Eq.~(\ref{eq:Dyson}) and the full procedure is iterated to self-consistency~\cite{Barbieri09a}.

\section{Calculations}
\label{sec:space}

Extremely large models spaces are not required for the present analysis because we already account for the short-range
part of the interactions through the partitioning procedure described in Sec.~\ref{sec:FRPA}~\cite{Barbieri09a}.
In the energy regime we are interested in, short-range physics affects mainly the
real part of the self-energy in the domain of interest. 
The contributions to the imaginary part are not included as they show up at very high
positive energies which are not considered here~\cite{Dickhoff04}.
The self-energies of $^{40}$Ca, $^{48}$Ca and  $^{60}$Ca were calculated using the FRPA
in a harmonic-oscillator model space with frequency $\hbar\Omega$ = 10~MeV.
Calculations for $^{60}$Ca were possible in no-core model spaces including up to 8 major shells
($N_{max} =$7) and we therefore employed this truncation for all the results presented 
in Sec.~\ref{sec:results}.  This space is deemed large enough to provide a proper description of the 
physics around the Fermi surface and qualitatively good at energies in the region of giant-resonance excitations which are of interest in this study.

Green's function theory---and in particular the FRPA---involves infinite summations of linked diagrams.
This implies that computational requirements scale favorably with the increase of the model-space size
and that the method is size extensive, which allows controlling theoretical errors when increasing
the size of the system.
The FRPA method has been tested in purely {\em ab-initio} calculations of $^4$He
in Ref.~\cite{Barbieri10var} and was found to achieve accuracies comparable to coupled-cluster results~\cite{Hagen07}. 
The further advantage of the FRPA formalism is that it calculates explicitly the effects of {\em all} many-body
excitations including the region of giant resonances. The result is a global description of the self-energy
over a wide range of energies. 
The FRPA is then the method of choice for our purpose of investigating medium-mass nuclei in a wide energy domain around the Fermi surface. 
We note that for a calculation of the ground-state energy the partition method implies that the contribution of high-momentum components still needs to be added~\cite{Muther95}.
Since these high-momentum components appear outside the energy domain of present interest, this issue is of no importance here.

%
\begin{figure}[t]
\includegraphics[width=0.9\columnwidth]{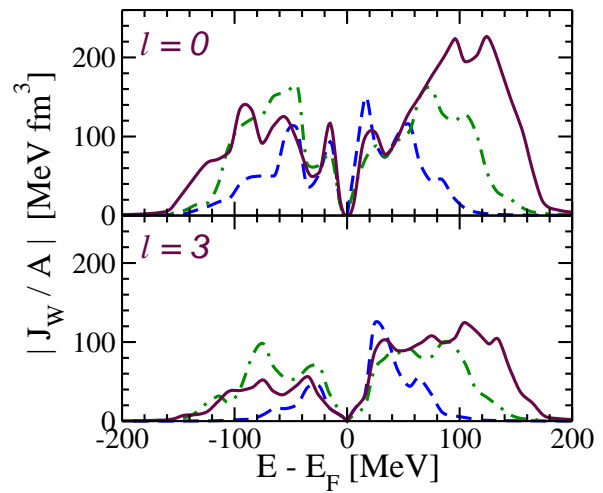}
\caption{Imaginary volume integral $J^\ell_W(E)$ of the $^{48}$Ca self-energy calculated with model spaces of different sizes.
The top~(bottom) panels refer to the scattering of a neutron with angular momentum $\ell$=1~($\ell$=3). Dashed, dot-dashed and
full lines refer to model spaces of 6, 8, and 10 oscillator shells, respectively. These results are for the N3LO interaction.
 }
\label{fig:ModSp_comp}
\end{figure}
%
In this work we will focus on averaged properties of the self-energy, as described by the volume
integrals~(\ref{eq:intgs}), for which meaningful results can be expected. 
These will be reliable within in a certain energy interval
due to the unavoidable truncation of the model space.
This window is centered around $E_F$ and increases with the size of the model space itself.
In order to assess these limits, we consider the $J_W$ of $^{48}$Ca obtained with the N3LO
interaction by Entem and Machleidt~\cite{Entem03} for model spaces of different sizes.
Calculations for this nucleus are possible including up to 10 major oscillator shells, as
reported in~\cite{Barbieri09b}.
Figure~\ref{fig:ModSp_comp} shows the proton $J_W(E)$ in the $\ell$=0 and $\ell$=3 partial waves
for models spaces of 6, 8, and 10 shells (and including all orbits with angular momentum up to $\ell \leq$7).
As expected, results are similar over a range of energies that increases with $N_{max}$. For higher positive energies
(in the particle scattering case) $J_W$ is expected to increase even further, however, the calculated values drop quickly
to zero due to the lack of degrees of freedom.
Based on Fig.~\ref{fig:ModSp_comp} one can expect that the self-energies calculated for $N_{max}$=7 (8 shells) and discussed in this work will be meaningful for energies in the range -100~MeV$<E-E_F<$100~MeV.

\begin{figure}[b]
\includegraphics[width=3.2in]{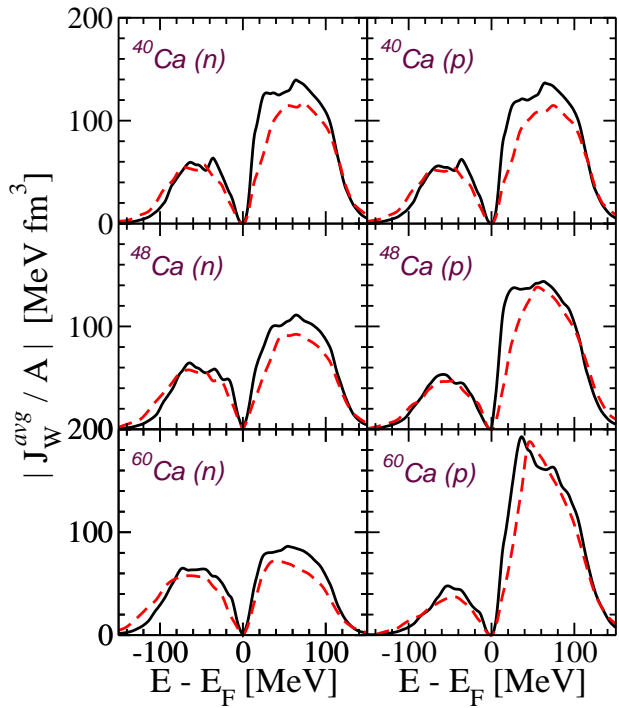}
\caption{ Comparison of the $J_W^{avg}$ for AV18 (solid line) and N3LO (dashed line).}
\label{fig:Jw_pots_comp}
\end{figure}
In addition to the chiral N3LO interaction, we have performed calculations with the realistic Argonne AV18 potential~\cite{Wiringa95}
which is representative of a local strongly-repulsive core.
The $J_W^{avg}$ for $^{40}$Ca, $^{48}$Ca, and $^{60}$Ca obtained with the two interactions are compared in Fig.~\ref{fig:Jw_pots_comp}.
The AV18  yields more absorption than the N3LO interaction, for $E > E_F$, especially in $^{40}$Ca where there is about 20\% more absorption. Below $E_F$ the absorption is only slightly higher. Another important difference is that the absorption strength of AV18 is enhanced at energies near $E_F$ and on the particle side. 
Nevertheless, it is clear from Fig.~\ref{fig:Jw_pots_comp} that the two interactions generate qualitatively similar results.
For this reason we will show mostly results for the AV18 in the following. 

\section{Results}
\label{sec:results}
\subsection{ Angular-Momentum Dependence }
\begin{figure*}
\includegraphics[width=1.4\columnwidth]{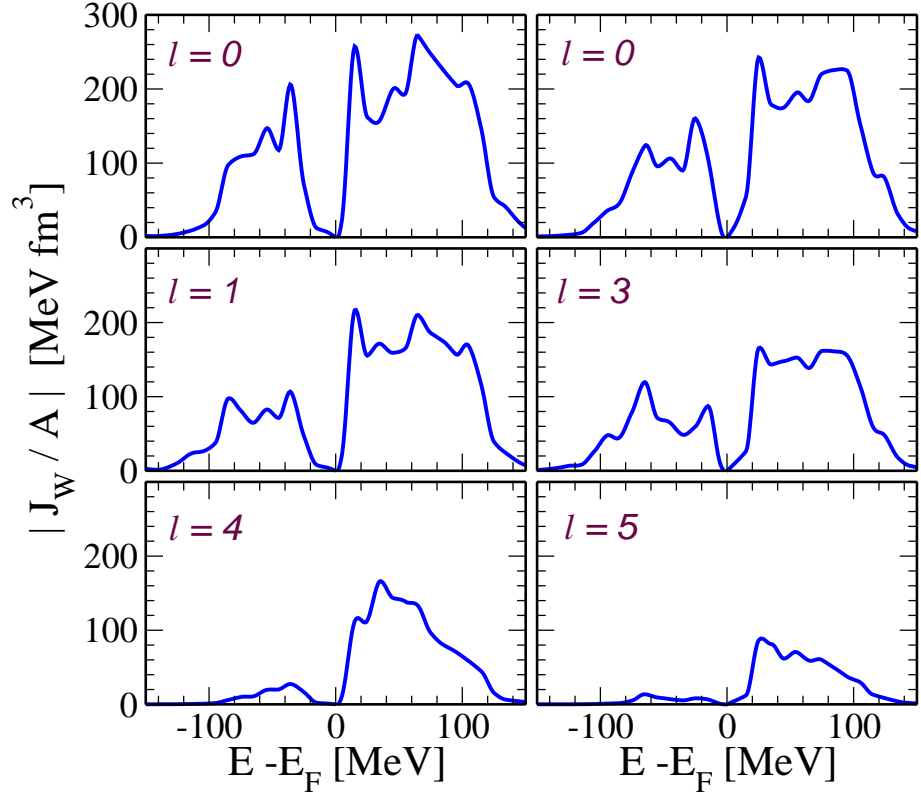}
\caption{ Imaginary volume integral $J_W^l$ of $^{40}$Ca self-energy for neutrons with $\ell = 0 - 5$.}
\label{fig:Jw_L_nca40}
\end{figure*}
\begin{figure*}
\includegraphics[width=1.4\columnwidth]{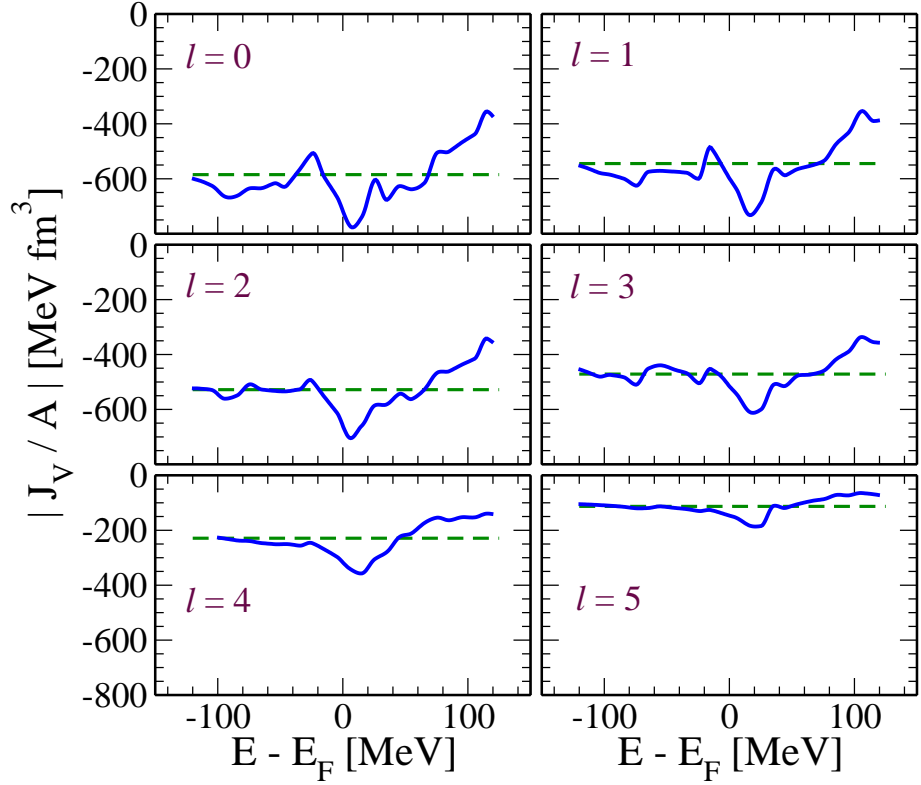}
\caption{ Volume Integrals of Re $\Sigma^{\ell}_0$ for neutrons in ${}^{40}$Ca. The horizontal, dashed lines are the volume integrals of $\Sigma^{\infty, \ell}_0(E_F)$. }
\label{fig:Jv_nca40}
\end{figure*}

Figures~\ref{fig:Jw_L_nca40} and~\ref{fig:Jv_nca40} 
give an overall example of the features of the imaginary ($J_W^\ell$) and real ($J_V^\ell$) 
part of the self-energy.  These results are shown for neutrons in ${}^{40}$Ca, employing the AV18 interaction and are separated in partial waves up to $\ell$=5.
In Fig.~\ref{fig:Jw_L_nca40}, plots of $J_W^\ell$ 
illustrate that absorption decreases systematically for increasing $\ell$ for neutrons, and naturally also for protons. 
As a consequence the variation of $J_V^\ell$ with respect to $J_F^\ell = J^\ell_V(E_F)$ obtained from $\Sigma^{\infty, \ell}_0(E_F)$, also decreases with increasing $\ell$ (Fig.~\ref{fig:Jv_nca40}) on account of the dispersion relation between the real and imaginary parts of the self-energy. 
The effect may be partly explained by the truncated model space, since the higher $\ell$-channels also have fewer orbits. 
On the other hand, the horizontal lines in Fig.~\ref{fig:Jv_nca40}, which are the contributions of $\Sigma^{\infty, \ell}_0$ to $J_F^\ell=J_V^\ell(E_F)$, clearly suggest that most of this decrease must arise from the $\ell$-dependence implied by the non-locality of the potential. 
Such an $\ell$-dependence suggests that it may be important to include non-local features in DOM potentials.
\begin{figure}[t]
\includegraphics[width=0.8\columnwidth]{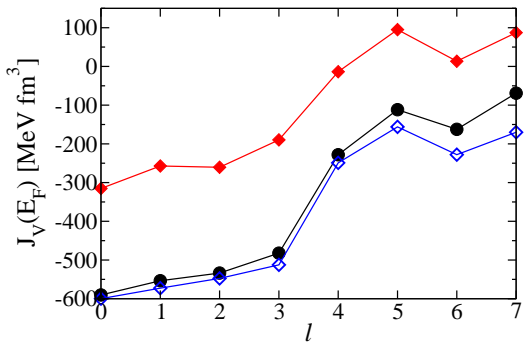}
\caption{
Angular momentum dependence for the volume Integrals $J_F^\ell = J^\ell_V(E_F)$ of $\Sigma^{\infty, \ell}(E_F)$ excluding the contribution of the dynamic part of the self-energy.
 For each $\ell$, results for protons are given by solid diamonds and neutrons by solid circles. Proton potentials are considerably less attractive due to the Coulomb energy. When the Coulomb interaction is suppressed (open diamonds) the proton results are close to the neutron results. The results shown are for $^{40}$Ca using the AV18 interaction.}
\label{fig:JvHF}
\end{figure}
For a given energy, both the static term and the dynamic term have similar radial shapes. In Fig.~\ref{fig:JvHF} the volume integrals $J_F^\ell = J^\ell_V(E_F)$ are shown excluding the contribution of the dynamic part. Note that because the proton potential is not as deep as that of the neutrons, the volume integral will be smaller for protons than for neutrons. When the calculation is done without the Coulomb potential, the volume integrals for the protons are comparable to those for the neutrons. 

\begin{figure}[b]
\includegraphics[width=0.75\columnwidth]{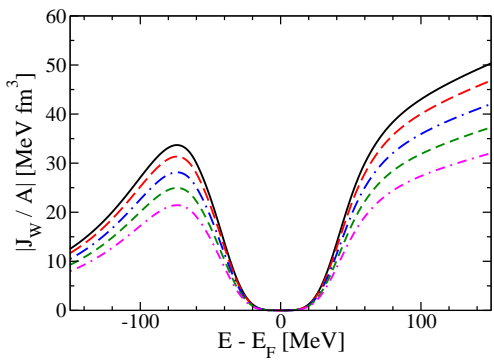}
\caption{ Imaginary volume integrals of the volume part of a DOM self-energy with a local Woods-Saxon form factor replaced by a non-local form proposed by Perey and Buck. The results shown are for $\ell = 0$ (solid), $\ell = 1$ (long-dash), $\ell = 2$ (long-dot-dash), $\ell = 3$ (short-dash) and $\ell = 4$ (short-dot-dash).  } 
\label{fig:Jw_nca40_dom}
\end{figure}
This effect of non-locality can be illustrated by taking \textit{e.g.} the energy dependence of the volume contribution of a DOM potential~\cite{charity07} and replacing the radial form factor by a non-local potential.
The radial parameters of such a non-local potential employed here correspond to the non-local Hartree-Fock potential of Ref.~\cite{Dickhoff11}.  Such a non-local potential is of the form proposed by Perey and Buck~\cite{Perey62} and contains a gaussian form factor describing the non-locality. The results are shown in Fig.~\ref{fig:Jw_nca40_dom}. 
Since the non-local potential depends on the angle between $\bm{r}$ and $\bm{r}'$ there is an automatic $\ell$-dependence of the projected $J_W^\ell$ that exhibit a systematic decrease in absorption for increasing $\ell$.
While it is apparently possible to fit elastic scattering data with local potentials, a non-local potential has a substantial effect on the interior scattering wave function and therefore \textit{e.g.} on the analysis of transfer reactions that rely on such wave functions~\cite{nunes11}.

\begin{figure}[t]
\includegraphics[width=2.7in]{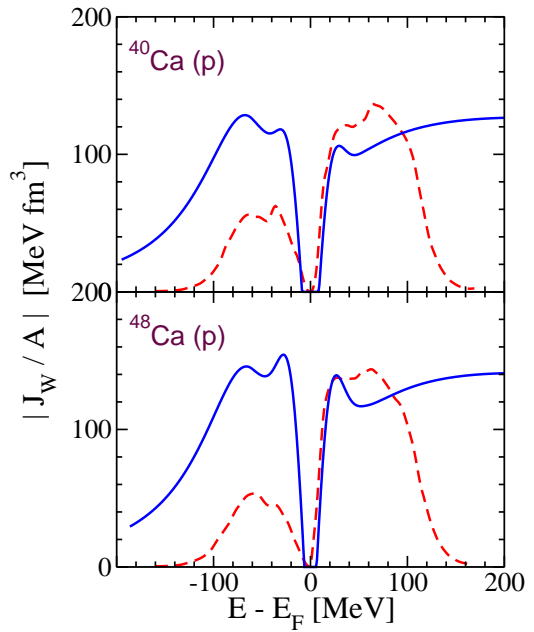}
\caption{The FRPA results for the average over all $\ell$-channels (dashed) are compared with the DOM result (solid), corrected for non-locality.}
\label{fig:pca40-48_vs_dom}
\end{figure}
The possible importance of non-locality for the calculation of observables below the Fermi energy was pointed out in Ref.~\cite{Dickhoff11}. 
When the real part of the self-energy at the Fermi energy is represented by a truly non-local potential, it becomes possible to properly calculate the spectral functions below the Fermi energy and observables like the charge density.
The importance of non-locality for the imaginary part of the self-energy suggested by the FRPA calculations may actually provide a handle on describing the nuclear charge density for ${}^{40}$Ca more accurately than was possible in Ref.~\cite{Dickhoff11}.

A direct comparison of $\ell$-averaged FRPA volume integrals with the corresponding DOM result is made in Fig.~\ref{fig:pca40-48_vs_dom}.
Since the DOM results are calculated from a local potential, they must be corrected by the effective mass that governs non-locality~\cite{Mahaux91,Dickhoff11}, before they can be compared with the FRPA results, which are generated from non-local potentials. The overall effect of this correction is to enhance the absorption. 
\begin{figure}[b]
\includegraphics[width=0.8\columnwidth]{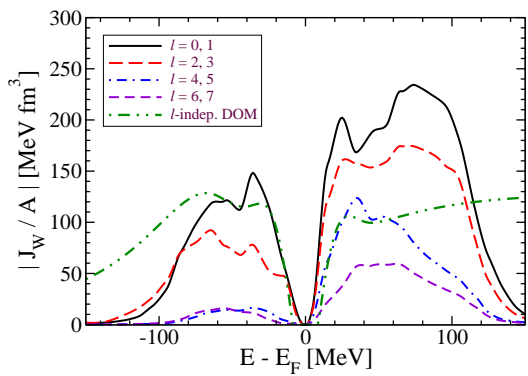}
\caption{Separate partial wave contributions of $J_W$ averaged over $\ell$-channels with the same number of harmonic-oscillator orbits in the model space. This plot is for neutrons in $^{40}$Ca. The dash-double-dotted curve represents the DOM result.}
\label{fig:nca40_avg_dom}
\end{figure}
Referring to Fig.~\ref{fig:pca40-48_vs_dom}, one can see that the FRPA exhibits different behavior above and below $E_F$ than is assumed in the DOM. The FRPA predicts that there is significantly less absorption below $E_F$ than above, whereas according to the assumptions made in a DOM fit, the absorption is roughly symmetric above and below up to about 50 MeV away from $E_F$~\cite{Mahaux91,charity06,charity07,Mueller11}. 
While this assumption is made in the local version of the DOM, the transition to a non-local implementation distorts this assumption of symmetry because the attendant correction involving the effective mass is different above and below the Fermi energy as can be seen in Fig.~\ref{fig:pca40-48_vs_dom}.
Since only the absorption above the Fermi energy is strongly constrained by elastic scattering data, it is encouraging that the $\ell$-averaged FRPA result is reasonably close to the DOM fit for both nuclei in the domain where the FRPA is expected to be relevant on account of the size of the chosen model space.
The simplifying assumptions of a symmetric absorption around $E_F$ and locality in the DOM generate unrealistic occupation of higher $\ell$-values below the Fermi energy which is not obtained in the FRPA. 
More insight into this result is obtained in Fig.~\ref{fig:nca40_avg_dom} where the volume integral is averaged over $\ell$-channels with the same number of harmonic-oscillator orbits inside the chosen model space for neutrons in ${}^{40}$Ca.
Below the Fermi energy the contribution with 3 and 4 orbits dominate associated with the prevalence of low-$\ell$ orbits like $s,p$, and $d$.
Higher $\ell$-values are less relevant below the Fermi energy and this is clearly illustrated in the figure.
The dash-double-dotted curve illustrates the DOM results also shown in Fig.~\ref{fig:pca40-48_vs_dom}.
The DOM result should therefore probably be compared below the Fermi energy with curves corresponding to the dominant $\ell$-values, whereas above the Fermi energy the higher $\ell$-values play a more prominent role.
Nevertheless, it is clear that the DOM overestimates the absorption of partial waves below the Fermi energy that are Pauli blocked in agreement with the observations in Ref.~\cite{Dickhoff11}.

Further comparison of FRPA with the DOM self-energy is made in Table I for the ph-gap.  The AV18 seems to provide smaller ph-gaps by 1-2 MeV compared to N3LO. However, in both cases these gaps substantially overestimate the experimental results (see Table I).
DOM fits from Ref.~\cite{charity07} are also included in the table and are typically closer to experiment.
\begin{table}[t]
\caption{Particle-Hole Gaps in MeV}
\begin{tabular}{|c|c|c|c|c|c|}
\hline\hline
\multicolumn{2}{|c|}{} & AV18 & N3L0 & DOM & Exp. \\ \hline
\multirow{2}{*}{$^{40}$Ca }& $\nu$ & 10.7 & 12.0 & 7.79 & 7.23 \\
& $\pi$ & 7.9 & 12.1 & 7.20 & 7.24 \\ \hline
\multirow{2}{*}{$^{48}$Ca} & $\nu$ & 4.8 & 4.9 & 2.83 & 4.79 \\
& $\pi$ & 11.6 & 13.5 & 6.78 & 6.18 \\ \hline
\multirow{2}{*}{$^{60}$Ca} & $\nu$ & 4.9 & 6.5 & 4.95 & - \\
& $\pi$ & 10.4 & 12.3 & 6.13 & - \\
\hline
\end{tabular}
\end{table}

\subsection{Parity Dependence}
\begin{figure}
\includegraphics[width=0.9\columnwidth]{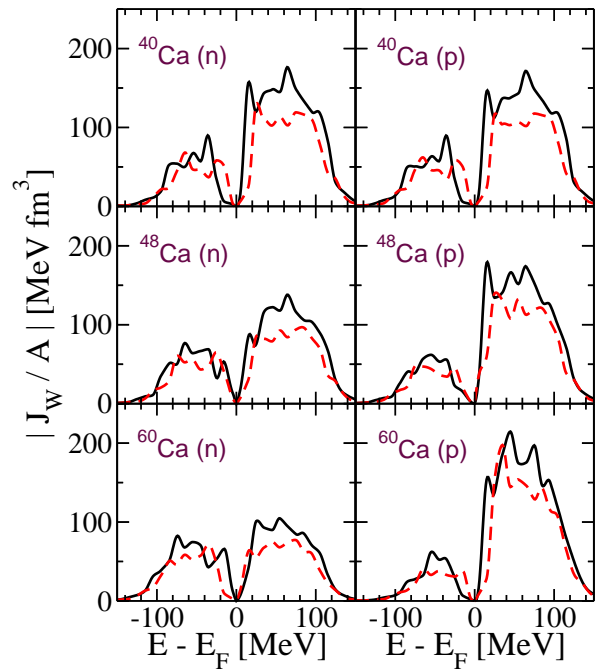}
\caption{ $J_W$ averaged over even $\ell$-channels (solid) is compared with Jw averaged over odd $\ell$-channels (dashed) } 
\label{fig:Jw_eve_odd}
\end{figure}

In Fig.~\ref{fig:Jw_eve_odd}, the absorption of the negative parity channels is compared with that of the positive parity channels in $^{40}$Ca, $^{48}$Ca, and $^{60}$Ca. The averages $\left(\sum_{\text{even } \ell}{J_W^\ell} \right)/ N_{\text{even } \ell}$ and $\left( \sum_{\text{odd } \ell}{J_W^\ell}\right) / N_{\text{odd } \ell}$ are compared in order to see the trends more clearly. 
An interesting feature in $^{40}$Ca is that just below $E_F$ there is more negative parity absorption than for even parity, while just above $E_F$ the opposite is true. 
The effect can be understood in terms of the number of 2p1h and 2h1p states, which are the configurations beyond the mean-field approximation that are closest to $E_F$. In these states, the ph and the hp phonons have negative parity, since the holes are in the $sd$-shell while the particles are in the $pf$-shell. Thus, near $E_F$, the 2h1p states will have negative parity and the 2p1h states will have positive parity. 

Proton ph-configurations at low energy continue to have negative parity,
as the neutron number increases in the $pf$-shell.
However, phonons with positive parity can be created at energies close to $E_F$ due to the partial filling of the neutron $pf$-shell.  
So, both parities for 2p1h and 2h1p states are possible. As a result, in $^{48}$Ca one sees little difference between the absorption from negative and positive parity states. 

In $^{60}$Ca, which is the next closed shell, the neutron $pf$-shell is filled and the corresponding low-lying neutron ph configurations again have negative parity, as in $^{40}$Ca; but in this case the neutron holes have negative parity corresponding to $\ell =1$ and 3. Thus, there are more 2h1p states with positive parity near $E_F$ for the neutrons. The situation for the protons is similar to the case of $^{40}$Ca. 
The inversion of the dominant parity above and below $E_F$ is quite general when major shells are filled or depleted and also visible in the partial waves separately.

\subsection{ Asymmetry Dependence}

\begin{figure*}
\includegraphics[width=1.8\columnwidth]{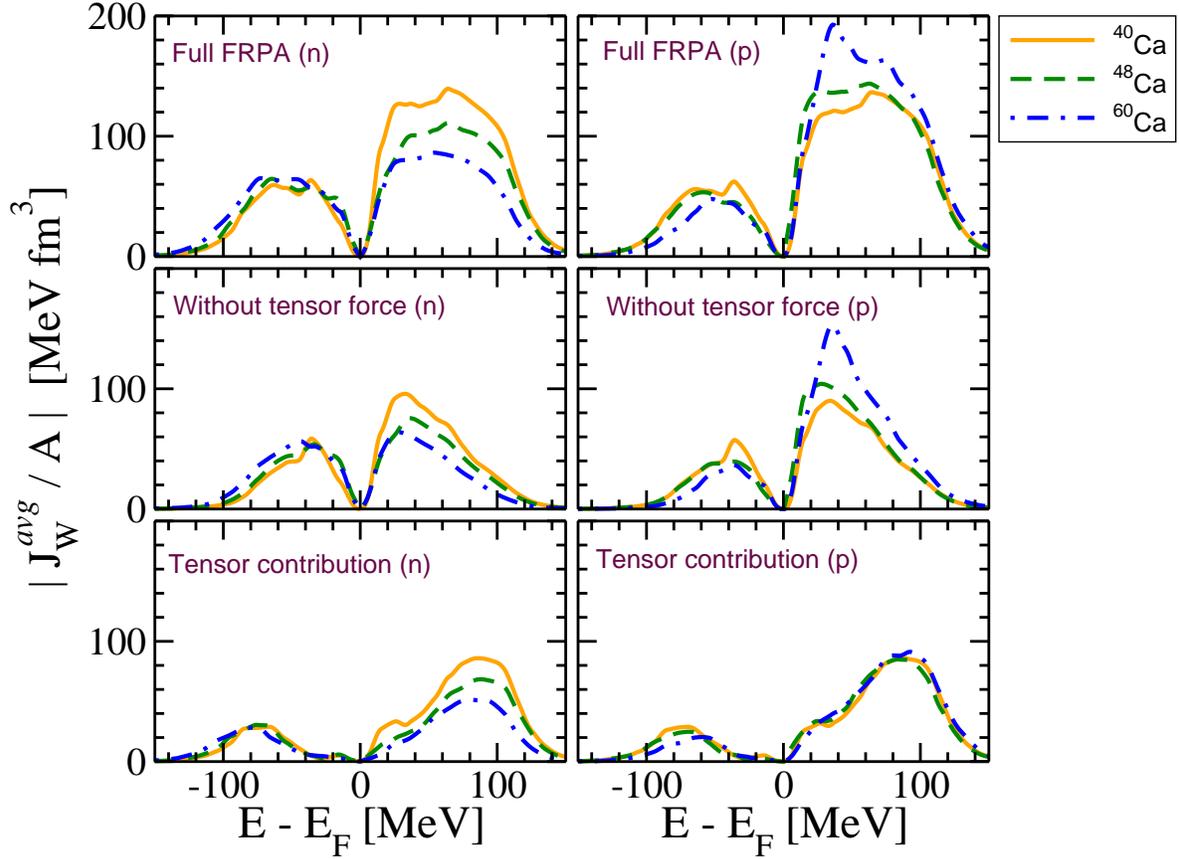}
\caption{Asymmetry dependence of the absorption for neutrons and protons and dependence on tensor correlations. The top panels shows $J_W^{avg}$ for $^{40}$Ca
(solid) is compared with the results for $^{48}$Ca (dashed), and $^{60}$Ca (dot-dashed). The middle  panels are obtained by suppressing the tensor component
in the AV18 interaction.
The difference between the top two panels is plotted in the bottom panels to provide a more detailed assessment of the correlations induced by the tensor force.}
\label{fig:Jw_asym_dep}
\end{figure*}

 The behavior of the nuclear self-energy with changing proton-neutron asymmetry ($\alpha = (N-Z)/A$) has important implications for unstable isotopes. Its understanding is fundamental in obtaining proper {\em global} parametrizations of the DOM so that these can be trusted in extrapolations toward the drip lines. Moreover, a strong absorption in the optical potential, even if at intermediate energies, affects the absolute quenching of spectroscopic factors~\cite{Barbieri09b}. Thus, the study of $J_W$ can in principle contribute to the much-debated asymmetry dependence of spectroscopic factors observed in knockout and transfer reactions~\cite{Gade08,Lee10,Timofeyuk09,Barbieri10var,Lee11,nunes11,Jensen11}. 

The $J_W^{avg}$  for the three different Ca isotopes are shown in the top panels of Fig.~\ref{fig:Jw_asym_dep}. These results predict an opposite behavior of the protons and neutrons above $E_F$, with the proton (neutron) potential increasing (decreasing) when more neutrons are added, qualitatively in agreement with expectations from the Lane potential model~\cite{Lane62}.
A recent DOM analysis based on several isotopes, including the Ca and Sn chains, employs a similar trend for the volume integrals~\cite{Mueller11}. However, the same analysis suggests different behavior of the imaginary surface contributions: neutron surface absorption appears to be rather independent of asymmetry, while variations are much stronger for protons and
for chains of isotopes tends to increase with asymmetry. 
The separation between volume and surface effects is an artifact of the functional form chosen for the optical model and such a separation cannot be carried out uniquely in a fully microscopic approach like  the present FRPA.
In general, one can argue that most of the physics at scattering energies below 50 MeV is dominated by surface effects which are well-covered by the FRPA, whereas volume effects pertain to higher energies, less well-covered by the FRPA chosen model space.
At energies below the Fermi surface, the overall absorption of both proton and neutron does not show much variations with changing asymmetry.
Since the DOM analysis employs less data from energies below $E_F$, this result must be further tested in future work.
Current DOM implementations assume that surface absorption is similar above and below the Fermi energy, which is clearly not suggested by the FRPA results.

The above pattern, in which one type of nucleon becomes more correlated when increasing the number of its isotopic partners, is a rather general feature in nuclear systems that is also found for asymmetric nucleonic matter~\cite{Frick05,Rios09}.  FRPA calculations of stable and drip-line nuclei show that this effect results in an asymmetry dependence of spectroscopic factors similar to that observed in knockout reactions, although the overall change from drip line to drip line is rather modest~\cite{Barbieri10var}.  We note, however, that there also exist other mechanisms that can affect this quenching besides the coupling to the giant resonance region, including a strong correlation to the ph gap~\cite{Barbieri09a} and effects of the continuum at the drip lines~\cite{Jensen11}.

\begin{figure}[t]
\includegraphics[width=0.85\columnwidth]{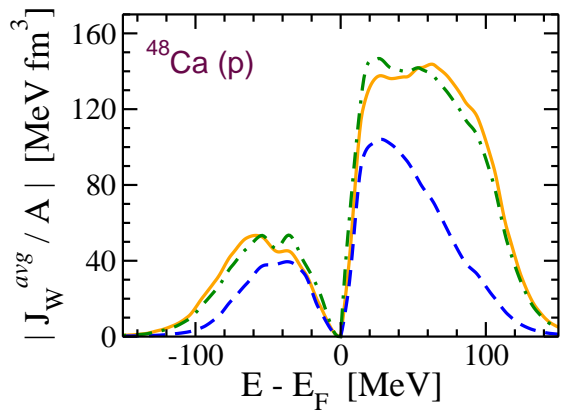}
\caption{ Effect of the tensor force and charge exchange on correlations on the proton-$^{48}$Ca self-energy. The solid curve is the imaginary volume integral $J_W^{avg}$ from the full FRPA calculation, while the dashed curve results from removing the tensor term in the AV18 interaction. The dash-dot curve is obtained by excluding charge exchange from the full calculation.  The same results are found for neutrons and the other Ca isotopes.}
\label{fig:tensor_chex}
\end{figure}

From the characteristics of the above asymmetry dependence one expects that the nuclear interaction between protons and neutrons plays a major role.  The tensor force of the nuclear interaction can provide one such mechanism since it is particularly strong in the $T=0$ sector. Moreover, it has already been shown to influence the evolution of single-particle energies at the Fermi surface~\cite{Otsuka05}.
To investigate its implication for single-particle properties at energies farther removed from $E_F$, we recalculated $J_W^{avg}$ by suppressing the tensor component of the AV18 interaction. This is shown in Fig.~\ref{fig:tensor_chex} for protons on $^{48}$Ca and in the middle panels of Fig.~\ref{fig:Jw_asym_dep} for all isotopes.
Its removal results in a drastic reduction of absorption at energies $|E-E_F|>$30~MeV. Thus, tensor effects give an important contribution to scattering at these energies.
%
The difference with the complete solution is plotted in the bottom panels of Fig.~\ref{fig:Jw_asym_dep} to highlight the separate effect of the tensor force although this is not unique due to the presence of interference terms.
It is apparent that the tensor force has a very significant effect on the correlations far from the fermi surface, but it contributes only to the
asymmetry dependence of neutron scattering. On the other hand, both scattering and negative energy states near the Fermi surface are dominated by correlations other than tensor which thus produce most of the asymmetry dependence obtained in the full calculation.

In Fig.~\ref{fig:tensor_chex} (dot-dashed line), we have also calculated the FRPA self-energy by suppressing charge-exchange excitations in the polarization propagator $\Pi(E)$. These contributions correspond to a mechanism in which the proton (neutron) projectile is Pauli exchanged with a neutron (proton) in the target and it was argued that it could enhance surface absorption due to the presence of Gamow-Teller resonances, with strength increasing as $\approx 3(N-Z)$~\cite{charity07}. However, the FRPA results suggest that charge-exchange excitations of the target interfere only very weakly with the nucleon-nucleus scattering process.





\section{Conclusions}
\label{sec:con}
In this investigation, an attempt was made to establish links between the DOM, an empirical approach to the nuclear-many body problem based on the framework of the Green's function method and relevant experimental data, and the microscopic FRPA approach. An analysis of the volume integrals calculated from both approaches proved to be a useful link, and on the whole, both the DOM and the FRPA produced similar results. However, there were some significant and illuminating differences. 

The FRPA exhibits some important shell effects as neutrons are added to $^{40}$Ca. In particular, there is a parity dependence in $^{40}$Ca and $^{60}$Ca, but not in $^{48}$Ca, where both parities occur at low energy due to the partial filling of the neutron $pf$-shell. Such an effect has not hitherto been taken into account in the DOM. Inspection of the imaginary volume integrals generated by the FRPA also calls into question the assumption in most DOM analyses that the imaginary part is symmetric about $E_F$ for the surface absorption.
Further insight into the underlying physics of DOM potentials is provided by the observation that a substantial contribution of the absorption is due to the NN tensor force.
This influence becomes dominant at energies about 40 MeV above or below $E_F$.
For protons, however, most of the observed asymmetry dependence of the absorption at positive energy in DOM fits appears to be due to central components of the interaction.
For neutrons the decrease in absorption at positive energy obtained with the FRPA must be contrasted with the weaker effects deduced so far from DOM fits.
Noteworthy is also the relevance of the non-locality of the absorption process obtained from the microscopic FRPA.
It leads to an important $\ell$-dependence that may play an important role in explaining data like the nuclear charge density that are associated with properties of the self-energy below the Fermi energy.
Its role in scattering processes remains to be studied as well and has important consequences for the analysis of transfer and knockout reactions which are sensitive to interior wave functions generated by optical potentials.

\acknowledgments This work was supported by the U.S.
National Science Foundation under grants PHY-0652900 and PHY-0968941,
by the Japanese Ministry of Education, Science
and Technology (MEXT) under KAKENHI grant no. 21740213, and by the United Kingdom Science and Technology Facilities Council (STFC) through travel grant No. ST/I003363.
%
%
SJW acknowledges support from the Japan-US JUSTIPEN program and the  
hospitality of the Theoretical Nuclear Physics Laboratory at RIKEN
(Japan) during the beginning of this work.

\bibliography{refs}
%
%
%
%

\end{document}